\documentclass[conference]{IEEEtran}
\usepackage{graphicx}
\usepackage{amsmath,epsfig,amssymb,verbatim,amsopn,cite,multirow}
\usepackage{amsmath}
\usepackage{bm}
\usepackage{algorithm}
\usepackage{algorithmic}
\usepackage{color}

\usepackage{cite}
\usepackage{url}
\usepackage{subfigure}
\usepackage[margin=14.5mm,top=17mm,bottom=24.5mm]{geometry}%margins;{geometry}%margins; tweak for printer deficiencies
\IEEEoverridecommandlockouts

\newcommand{\proof}{\noindent\textit{\textbf{Proof.}} }
\newcommand{\proofend}{\hfill$\square$}

\newenvironment{Proof}
  {\proof}{\proofend}
\newtheorem{proposition}{Proposition}

\newcommand{\qq}{{\bf q}}

\newcommand{\qt}{{\bf t}}

\begin{document}
\bstctlcite{IEEEexample:BSTcontrol}  
\title{Delay Alignment Modulation for Secure \\ ISAC Systems}
\author{
    \IEEEauthorblockN{Tianyu Lu, Jiajun He, Mohammadali Mohammadi, and Michail Matthaiou}
    \IEEEauthorblockA{
         Centre for Wireless Innovation (CWI), Queen’s University Belfast, Belfast, U.K. \\
        E-mail: \{t.lu, j.he, m.mohammadi, m.matthaiou\}@qub.ac.uk}
        \thanks{This work was supported by the European Research Council (ERC) under the European Union’s Horizon 2020 research and innovation programme (grant agreement No. 101001331) and by a research grant from the Department for the Economy Northern Ireland under the US-Ireland R\&D Partnership Programme. It is also based upon work from COST Action 6GPHYSEC (CA22168), supported by COST (European Cooperation in Science and Technology).}
        }


\markboth{Journal of \LaTeX\ Class Files,~Vol.~X, No.~X, X~X}%
{Shell \MakeLowercase{\textit{et al.}}: Bare Demo of IEEEtran.cls for IEEE Journals}



 \maketitle

\begin{abstract}
This paper introduces delay-alignment modulation (DAM) for secure integrated sensing and communication (ISAC). Due to the broadcast nature of multi-user downlinks, communications are vulnerable to eavesdropping. DAM applies controlled per-path symbol delays at the transmitter to coherently align the multipath components at the intended user, enhancing the received signal power, while simultaneously creating delay misalignment at the eavesdropper (Eve). To mitigate sensing degradation caused by multipath propagation, we propose a two-stage protocol that first estimates the angle and then the delay of the line-of-sight (LoS) path after suppressing multipath interference. We derive the secrecy spectral efficiency (SSE) and the Cramér–Rao bound (CRB) of the target delay. Finally, we develop a path-based zero-forcing (ZF) precoding framework and formulate a max–min SSE design under CRB and power constraints. Simulation results show DAM significantly outperforms the strongest-path (SP) benchmark in terms of SSE, while meeting sensing requirements, since intentional delay alignment at legitimate users degrades Eve’s reception.
\end{abstract}

\vspace{0.4em}

\IEEEpeerreviewmaketitle
\section{Introduction}
\IEEEPARstart{I}{SAC} offers a unified framework, where a single waveform supports both data transmission and radar sensing—an approach seen as key to mitigating spectrum scarcity in 6G networks~\cite{Liu2022}. However, the broadcast nature of ISAC downlink signals exposes them to eavesdropping, making secure waveform and precoding design a critical challenge~\cite{Zhu2025}.

A growing body of research has focused on enhancing physical-layer security (PLS) in ISAC systems~\cite{Chu2023,Chu2024,Su2023}. 
In~\cite{Chu2023}, the authors proposed a joint optimization of communication and radar transmit precoders to minimize the worst-case eavesdropping signal-to-interference-plus-noise ratio (SINR) across multiple legitimate users, under the assumption that Eve has channel state information (CSI). In~\cite{Chu2024}, the authors introduced the use of reconfigurable intelligent surfaces in ISAC systems to enhance secrecy by dynamically shaping the wireless environment, while artificial noise (AN)-aided precoding was proposed in~\cite{Su2023} to suppress interception when channel disparity is limited. Although most studies adopt narrowband models, frequency-division multiplexing (OFDM)-based ISAC has recently gained attention~\cite{Yang2023,Han2025}, with resource allocation optimized for secrecy rate under sensing constraints. However, OFDM’s high peak-to-average power ratio (PAPR) poses challenges~\cite{Xiao2023}, motivating the search for alternative waveform designs.

DAM is a single-carrier technique that relaxes the high PAPR and strict orthogonality constraints of multicarrier waveforms~\cite{Lu_DDAM}. By pre-compensating delays and applying path-specific precoding, DAM aligns multipath components at the receiver to eliminate intersymbol interference (ISI) and enhance signal strength. Beyond ISI removal, the user-tailored delay/angle alignment creates temporal–spatial mismatches at unintended locations, reducing Eve’s effective signal power. Hence, unlike AN-based designs that divert transmit power to AN, DAM offers “built-in” interference to the Eve without extra power expenditure, while preserving the desired user’s strength. Motivated by these observations, this paper investigates DAM-assisted secure ISAC. We consider a multi-user downlink ISAC system, in which the base station (BS) senses a target while an Eve attempts to intercept all users in multi-path environments. Our main contributions are summarized as follows:
\vspace{-0.2em}
\begin{itemize}
\item We propose a two-stage ISAC protocol: first, the BS estimates the angles and delays of the target and scatterers; then, ZF precoding suppresses the multipath components, preserving the LoS path for refined delay estimation and DAM-assisted secure multiuser communication.
%We propose a two-stage ISAC protocol: first, the BS estimates angles and delays of the target and scatterers; then, ZF precoding suppresses the dominant paths, enabling refined delay estimation and DAM-assisted secure multiuser communication.
%We propose a two-stage ISAC protocol. In the first stage, the BS senses and estimates the angles and delays of the target and surrounding scatterers. In the second stage, based on the estimated multipath information, ZF precoding is employed to suppress the dominant paths, enabling refined delay estimation and simultaneous DAM-assisted secure communication for multiple users.
\item We derive closed-form expressions for the SSE and the CRB to jointly characterize the communication security and sensing accuracy. A path-based ZF precoding framework is developed to mitigate ISI, inter-user interference (IUI), and sensing interference. The precoders are further optimized to maximize the worst-user SSE under CRB and power constraints.
\item We benchmark DAM against the SP scheme. Simulation results show that DAM consistently achieves higher SSE than SP, while satisfying sensing requirements. Moreover, the root mean squared error (RMSE) of the proposed delay estimator closely approaches the CRB, confirming the tightness of the analytical bound.
\end{itemize}

\emph{Notations:} Boldface lowercase letters and boldface uppercase letters denote vectors and matrices, respectively; $\operatorname{diag}(\cdot)$ forms a diagonal matrix out of its vector argument; $\operatorname{vecd}(\cdot)$ forms a vector out of the diagonal of its matrix argument; $\operatorname{vec}(\cdot)$ vectorizes its matrix argument by stacking its columns into a single column vector; $\operatorname{blkdiag}(\cdot)$ denotes the block-diagonal operator placing its arguments along the main diagonal; $(\cdot)^T$, $(\cdot)^H$, $(\cdot)^*$, $(\cdot)^{-1}$ denote the transpose, conjugate transpose, conjugate, and inverse, respectively; %$\mathbb{C}^{m\times n}$ is the complex space of a $m\times n$ matrix. 
$\mathbb{E}\{\cdot\}$ denotes the statistical expectation; $\diamond$ denotes the Khatri–Rao matrix product; $\odot$ denotes the convolution operation; $\otimes$ denotes the Kronecker product; $[x]^+=\max\{x,0\}$; %The discrete-time impulse function 
$\delta[n-k]$ equals $1$ when $n=k$ and $0$ otherwise. Finally,  $\Vert\cdot\Vert_2$ denotes the Euclidean norm.

\vspace{-0.1em}
\section{System Model}

%\subsection{System Overview}
We consider a monostatic ISAC system. %as shown in Fig.~\ref{System_Model}. 
The BS employs a uniform linear array (ULA) consisting of $N$ transmit antennas with inter-element spacing $d$. In addition, a dedicated single receive antenna is placed adjacent to the ULA to collect echo signals for sensing. The BS simultaneously serves $K$ single-antenna user equipment (UEs) and probes a target, in the presence of a single-antenna Eve. The system operates over a wideband, frequency-selective channel, and transmission is carried out using DAM. The proposed protocol operates in two stages. In Stage~1, the sensing module estimates the target’s angle in a multipath environment (Section~\ref{sec:angle_sensing}). In Stage~2, leveraging the angle estimate, the BS designs sensing precoding to suppress multipath interference and to refine the delay estimate of the target (Section~\ref{sec:delay_sensing}). In parallel, the communication module applies DAM-based precoding to deliver confidential data to UEs while suppressing information leakage to Eve (Section~\ref{sec:secure_communication}). 
%Because sensing and communication share the same antenna aperture and time–frequency resources, we further design a joint precoding strategy that balances sensing accuracy and secrecy performance.
% %%%%%%%%%%%%%%%%%%%%%%%%%%%%%%%%
% \begin{figure}[t!]
%   \centering
%   \includegraphics[width=2.8in]{System_Model2.pdf}
%   \vspace{-0.2em}
%   \caption{Illustration of the secure ISAC system (the orange antenna is dedicated to sensing).}
%   \label{System_Model}
%   \vspace{0.4em}
% \end{figure}
% %%%%%%%%%%%%%%%%%%%%%%%%%%%%%%%%%%%

\vspace{-0.3em}
\subsection{Propagation Environment} 
The system is defined in a two-dimensional (2-D) Cartesian coordinate system, with the ULA aligned along the $y$-axis. The UEs, target, and Eve are located on the positive $y$-axis. We consider both LoS and non-line-of-sight (NLoS) propagation between the BS and each node. All scatterers are assumed to lie on the horizontal plane. When a plane wave impinges on the BS from azimuth angle $\varphi$, the BS array response vector is given by $\mathbf{a}(\varphi) = \frac{1}{\sqrt{N}}[1,\dots,e^{j2\pi(N-1)d\sin{\varphi}/\lambda}]^T$, where $\lambda$ is the carrier wavelength.

\subsubsection{Sensing Channel} 
We consider a quasi-static block-fading model, where the channel remains constant within each coherence interval $T_c$. Given the system bandwidth $B$, the sampling period is defined as $T = 1/B$. The round-trip sensing channel between the BS and the target is assumed to consist of $L_s$ resolvable propagation paths. At the $n$-th sampling instant, the discrete-time sensing channel can be expressed as $\mathbf{h}_{s}^H[n] = \sum\nolimits_{l=1}^{L_s} \mathbf{h}_{s,l}^H \, \delta[n - n_{s,l}]$, where $\mathbf{h}_{s,l} \in \mathbb{C}^{N \times 1}$ denotes the channel vector associated with the $l$-th path, $n_{s,l} = B(\tau_{s,l} - \eta_s)$ is the corresponding discrete delay (assumed to be integer-valued), $\tau_{s,l}$ is the physical propagation delay, and $\eta_s$ denotes the reference timing corresponding to the earliest arriving path. 
%\comm{where $k = B(\tau - \eta_s)$ is the discrete delay corresponding to a continuous-time delay $\tau$.}\com{Already defined in 3-4 lines above.}

The $l$-th sensing path can be further modeled as $\mathbf{h}_{s,l} = \beta_{s,l}\mathbf{a}(\phi_{s,l})$, where $\beta_{s,l} = \sqrt{\frac{N}{L_s}}\alpha_{s,l}$ denotes the complex path coefficient, %incorporating large-scale and small-scale fading, $\alpha_{s,l}$ is the complex gain, 
while $\phi_{s,l}$ represents the azimuth angle. %of the $l$-th path (either LoS or NLoS).

\subsubsection{Communication Channel}
%In a wideband multipath environment, the discrete-time multiple-input single-output (MISO) 
The channel between the BS and the $k$-th UE is modeled as $\mathbf{h}_{c,k}[n] = \sum\nolimits_{l=1}^{L_{c,k}} \mathbf{h}_{c,kl} \, \delta[n - n_{c,kl}]$, where $\mathbf{h}_{c,kl}\!\in\!\mathbb{C}^{N\times 1}$ denotes the spatial channel vector of the $l$-th path, $L_{c,k}$ is the number of resolvable paths, $n_{c,kl} = B(\tau_{c,kl} - \eta_{c,k})$ is the discrete-time delay, $\tau_{c,kl}$ is the physical propagation delay, and $\eta_{c,k}$ denotes the timing reference of the $k$-th UE. The $l$-th path of the $k$-th UE is given by $\mathbf{h}_{c,kl} = \beta_{c,kl} \, \mathbf{a}(\phi_{c,kl})$, where $\beta_{c,kl} = \sqrt{\frac{N}{L_{c,k}}} \, \alpha_{c,kl}$, $\alpha_{c,kl}$ is the complex gain, and $\phi_{c,kl}$ denotes the azimuth angle.

Similarly, the eavesdropping channel is modeled as $\mathbf{h}_{e}[n] = \sum\nolimits_{l=1}^{L_e} \mathbf{h}_{e,l} \, \delta[n - n_{e,l}]$, where $\mathbf{h}_{e,l}\!\in\!\mathbb{C}^{N\times 1}$ denotes the spatial channel vector of the $l$-th path, $L_e$ is the number of resolvable paths, and $n_{e,l}$ is the discrete-time delay.

\vspace{-0.3em}
\subsection{Signal Model}

\subsubsection{Sensing Signal} 
The sensing waveform employs a unit-energy sinc pulse as the transmit shaping filter. 
Let $\{s_d[n]\}$ denote the discrete-time sensing sequence in the complex baseband domain, where $s_d[n]$ is an independent and identically distributed (i.i.d.) unit-power probing symbol, i.e., $\mathbb{E}\{|s_d[n]|^2\}=1$. With the sensing precoder $\mathbf{f}_s \in \mathbb{C}^{N\times1}$, the transmit vector for sensing is expressed as $\mathbf{x}_{s}[n] = \mathbf{f}_{s}s_d[n]$. The average transmit power of the sensing waveform satisfies $\mathbb{E}\{\|\mathbf{x}_s[n]\|_2^2\} = \|\mathbf{f}_s\|_2^2 = P_s$, 
where $P_s$ denotes the allocated sensing power.

\subsubsection{Communication Signal}
The downlink communication employs DAM. 
Let $s_{c,k}[n]$ denote the discrete-time data symbol sequence for the $k$-th UE, where $s_{c,k}[n]$ are i.i.d. zero-mean and unit-power symbols, i.e., $\mathbb{E}\{|s_{c,k}[n]|^2\}\!=\!1$. 
The per-UE transmit signal under DAM is given by $\mathbf{x}_{c,k}[n] \!=\! \sum\nolimits_{l=1}^{L_{c,k}} \mathbf{f}_{c,kl}s_{c,k}[n \!-\! \kappa_{c,kl}]$, where $\mathbf{f}_{c,kl}\in\mathbb{C}^{N\times1}$ denotes the precoding vector associated with the $l$-th path of the $k$-th UE, and $\kappa_{c,kl}$ is the pre-applied discrete delay used for delay alignment. 
The average transmit power of UE~$k$ satisfies 
$\mathbb{E}\{\|\mathbf{x}_{c,k}[n]\|_2^2\} = \sum_{l=1}^{L_{c,k}}\|\mathbf{f}_{c,kl}\|_2^2$, 
and the per-UE power constraint is enforced as
$\sum_{l=1}^{L_{c,k}}\|\mathbf{f}_{c,kl}\|_2^2 \!= \!P_{c,k}$, 
where $P_{c,k}$ denotes the transmit power for UE~$k$.

%%%%%%%%%%%%%%%%%%%%%%%%%%%%%%%%
\section{Target Sensing}
%%%%%%%%%%%%%%%%%%%%%%%%%%%%%%%%%
In the first stage, the BS sends probing signals, gathers echoes from the target and scatterers, and estimates angles and delays. The resulting spatial information supports later precoding refinement and secure transmission design. In the second stage, using these estimates, ZF precoding suppresses NLoS interference and enables more accurate delay estimation on the LoS paths.

%----------------------
\vspace{-0.5em}
\subsection{Angle Sensing}~\label{sec:angle_sensing}
%-------------------------
\vspace{-1.5em}
\subsubsection{Discrete-Time Received Signal}
%--------------------------
We define $h_s[m]$ as the finite impulse response (FIR) filter that characterizes the wideband sensing channel in the time domain. The FIR filter consists of $M$ taps determined by the multipath propagation and the sensing precoder $\mathbf{f}_s$. The $m$-th tap is defined as $$h_s[m] = \mathbf{h}^H_s[m]\mathbf{f}_s = \sum\nolimits_{l=1}^{L_s}\beta_{s,l}^*\mathbf{a}^H(\varphi_{s,l})\mathbf{f}_s\,\delta[m - n_{s,l}].$$ For compactness, $h_s[m]$ can be rewritten in matrix form as $h_s[m] = \mathbf{q}^H[m]\mathbf{D}_s^*\mathbf{A}_s^H\mathbf{f}_s$, where $\mathbf{A}_s = [\mathbf{a}(\varphi_{s,1}),\dots,\mathbf{a}(\varphi_{s,L_s})]\!\in\!\mathbb{C}^{N\times L_s}$ is the array steering matrix,  
$\mathbf{D}_s = \mathrm{diag}\{\beta_{s,1},\dots,\beta_{s,L_s}\}\!\in\!\mathbb{C}^{L_s\times L_s}$ contains the path gains, and $\mathbf{q}[m] = [\delta(m-n_{s,1}),\dots,\delta(m-n_{s,L_s})]^T\!\in\!\mathbb{C}^{L_s\times1}$ characterizes the time-domain structure.

At the $n$-th time instant, the discrete-time received signal is 
%---------------
\begin{align}
   y_s[n] = \mathbf{s}_d^H[n]\mathbf{h}_s + w_s[n],  
\end{align}
%--------------
where $w_s[n]$ is the additive white Gaussian noise (AWGN) with variance $\sigma_d^2$; 
$\mathbf{s}_d[n]=[s_d^*[n],\dots,s_d^*[n-M+1]]^T\!\in\!\mathbb{C}^{M\times1}$; $\mathbf{h}_s=[h_s[0],\dots,h_s[M-1]]^T\!\in\!\mathbb{C}^{M\times1}$ is the delay-domain channel after precoding, which is expressed as $\mathbf{h}_s = \mathbf{H}_s^H\mathbf{f}_s = \mathbf{G}_s^H\mathbf{D}_s^*\mathbf{A}_s^H\mathbf{f}_s$, with $\mathbf{H}_s \triangleq \mathbf{A}_s\mathbf{D}_s\mathbf{G}_{s}\in\mathbb{C}^{N\times M}$ is the time-space channel matrix and $\mathbf{G}_s=[\mathbf{q}[0],\dots,\mathbf{q}[M-1]]\!\in\!\mathbb{C}^{L_s\times M}$.

\subsubsection{Slot Representation}
To recover the delay-domain channel, we stack $M$ consecutive time samples into the $q$-th slot vector
%--------------
\vspace{-0.2em}
\begin{align}
   \mathbf{y}_{s,q} = \mathbf{S}_a^H\mathbf{H}_{s}^H\mathbf{f}_s + \mathbf{w}_{s,q}, 
\end{align}
%---------
where $\mathbf{S}_a = [\mathbf{s}_d[n],\dots,\mathbf{s}_d[n+M-1]] \in\mathbb{C}^{ M\times M}$ and $\mathbf{w}_{s,q} = [w_s[n],\dots,w_s[n+M-1]]^T \in\mathbb{C}^{ M\times 1}$ is the AWGN vector.

\subsubsection{Subframe Representation}
To recover the channel in the spatial domain, we stack $N$ consecutive slots into the $i$-th subframe, 
where each slot is transmitted using a distinct sensing precoder. The resulting subframe observation is expressed as 
%-------------
\begin{align}
   \mathbf{Y}_{s,i} = [\mathbf{y}_{s,1},\dots,\mathbf{y}_{s,N}] = \mathbf{S}_a^H\mathbf{H}_s^H\mathbf{F}_s + \mathbf{W}_{s,i}, 
\end{align}
%-------------
where $\mathbf{Y}_{s,i}\in\mathbb{C}^{M\times N}$, $\mathbf{F}_s = [\mathbf{f}_{s,1},\dots,\mathbf{f}_{s,N}]\in\mathbb{C}^{N\times N}$ and $\mathbf{W}_{s,i} = [\mathbf{w}_{s,1},\dots,\mathbf{w}_{s,N}]\in\mathbb{C}^{M\times N}$. 

By left/right inversion of the probing and precoding operators, an estimate of $\mathbf{H}_s$ is obtained as $\widehat{\mathbf{H}}_s^H = \mathbf{H}_s^H + \widehat{\mathbf{W}}_{s,i}$, where $\widehat{\mathbf{W}}_{s,i} = \widetilde{\mathbf{S}}_a\mathbf{W}_{s,i}\widetilde{\mathbf{F}}_s\in\mathbb{C}^{M\times N}$, with $\widetilde{\mathbf{S}}_a = (\mathbf{S}_a\mathbf{S}_a^H)^{-1}\mathbf{S}_a\in\mathbb{C}^{M\times M}$ and $\widetilde{\mathbf{F}}_s = \mathbf{F}_s^H(\mathbf{F}_s\mathbf{F}_s^H)^{-1}\in\mathbb{C}^{N\times N}$. 

We vectorize the estimated time–space channel matrix $\widehat{\mathbf{H}}_s$ as 
%---------------
\begin{align}
   \widehat{\mathbf{h}}_{s,j} = \operatorname{vec}(\mathbf{H}_s) + \widehat{\mathbf{x}}_{s,j} 
                               \overset{(a)}{=} (\mathbf{G}_{s}^T\diamond\mathbf{A}_s)\mathbf{a} + \widehat{\mathbf{x}}_{s,j},
\end{align}
%--------------
where $\widehat{\mathbf{h}}_{s,j}\in\mathbb{C}^{MN\times 1}$, $\widehat{\mathbf{x}}_{s,j} = \operatorname{vec}(\widehat{\mathbf{W}}_{s,j}^H)\in\mathbb{C}^{NM\times 1}$, and $\mathbf{a} = \operatorname{vecd}(\mathbf{D}_s) \in\mathbb{C}^{L_s\times 1}$. The equality $(a)$ follows from the identity 
$\operatorname{vec}(\mathbf{A}\mathbf{B}\mathbf{C})=(\mathbf{C}^T\diamond\mathbf{A})\operatorname{vecd}(\mathbf{B})$ 
when $\mathbf{B}$ is diagonal. Moreover, since $\operatorname{vec}(\mathbf{A}\mathbf{B}\mathbf{C})=(\mathbf{C}^T\otimes\mathbf{A})\operatorname{vec}(\mathbf{B})$, 
the processed noise term can be expressed as $\widehat{\mathbf{x}}_{s,j} = (\widetilde{\mathbf{S}}_a^* \otimes \widetilde{\mathbf{F}}_s^H )\mathbf{x}_{s,j}$, where $\mathbf{x}_{s,j} = \operatorname{vec}(\mathbf{W}^H_{s,j})\in\mathbb{C}^{MN\times 1}$. 

We define $\mathbf{U}\in\mathbb{C}^{MN \times L_s}$ as $\mathbf{U} = \mathbf{G}_{s}^T\diamond\mathbf{A}_s = \big[\boldsymbol{u}(\varphi_{s,1}, \tau_{s,1}),\dots,\boldsymbol{u}(\varphi_{s,L_s}, \tau_{s,L_s})\big]$. We have $\mathbf{G}_{s}^T = [\mathbf{p}(\tau_{s,1}),\dots,\mathbf{p}(\tau_{s,L_s})]\in\mathbb{C}^{M \times L_s}$, where $\mathbf{p}(\tau_{s,l}) = [\delta(0 + B(\eta_s - \tau_{s,l})),\dots,\delta(M-1 + B(\eta_s - \tau_{s,l}))]^T\in\mathbb{C}^{M \times 1}$. Hence, $\boldsymbol{u}(\varphi_{s,l}, \tau_{s,l}) = \mathbf{p}(\tau_{s,l})\otimes\mathbf{a}(\varphi_{s,l})$ is an $MN \times 1$ vector called as the space-time array manifold.

\subsubsection{Frame Representation} 
To enable subspace-based estimation, we stack multiple subframes to form a sensing frame. We define the channel gain at the $j$-th subframe as $\mathbf{a}_j$. By stacking the $Q$ subframes, we obtain $\widehat{\mathbf{H}}_Q = [\widehat{\mathbf{h}}_{s,1},\dots,\widehat{\mathbf{h}}_{s,Q}] = \mathbf{U}\mathbf{A}_Q + \widehat{\mathbf{X}}_Q$, where $\widehat{\mathbf{H}}_Q\in\mathbb{C}^{NM\times Q}$, $\mathbf{A}_Q = [\mathbf{a}_1,\dots,\mathbf{a}_Q]\in\mathbb{C}^{L_s\times Q}$, and $\widehat{\mathbf{X}}_Q = [\widehat{\mathbf{x}}_{s,1},\dots,\widehat{\mathbf{x}}_{s,Q}]\in\mathbb{C}^{NM\times Q}$. With this frame model, we can form the sample covariance 
$\widehat{\mathbf{R}}_Q = \widehat{\mathbf{H}}_{Q}\widehat{\mathbf{H}}_{Q}^{H}/Q\!\in\!\mathbb{C}^{MN\times MN}$
and apply subspace methods (e.g., 2D-MUSIC) to estimate the $2L_s$ structural parameters of interest 
$\{(\varphi_{s,l},\tau_{s,l})\}$.

%\subsection{ZF precoding}

\vspace{-0.2em}
\subsection{Delay Sensing}~\label{sec:delay_sensing}
Using the angles and delays estimated in the first stage, we apply ZF precoding to suppress interference from the NLoS components and refine delay estimation on the LoS paths. Meanwhile, secure downlink communication proceeds.
\subsubsection{ZF Precoding for Delay Sensing}

The received signal at the BS through the sensing channel is given by $y_{s}[n] = \mathbf{h}_{s}^H[n]\odot\mathbf{x}_s[n] + \mathbf{h}_{s}^H[n]\odot\mathbf{x}_c[n] + z_s[n]$, which can be expanded as 
%-------------
\vspace{0.1em}    
\begin{align}~\label{eq:received_signal_delay_sensing1}
        &y_{s}[n] = \mathbf{h}_{s,1}^H\mathbf{f}_{s}s_d[n - n_{s,1}] + \sum\nolimits_{l\neq1}^{L_s}\mathbf{h}_{s,l}^H\mathbf{f}_{s}s_d[n - n_{s,l}] + 
        \nonumber\\
        &\hspace{-1em}\sum\nolimits_{l=1}^{L_s}\!\!\sum\nolimits_{k=1}^{K}\!\!\sum\nolimits_{l^{\prime}=1}^{L_{c,k}}\!\mathbf{h}_{s,l}^H\mathbf{f}_{c,kl^{\prime}}s_{c,k}[n\!-\!\kappa_{c,kl^{\prime}}\!-\!n_{s,l}] \!+\!\! z_s[n],
    \end{align}
%----------
where $z_s[n]$ is the sensing noise with variance $\sigma^2_a$. With the precoding design for $\mathbf{f}_{c,kl^{\prime}}$ (c.f. Section \ref{sec: ZF_Optimized}), the communications interference in~\eqref{eq:received_signal_delay_sensing1} can be mitigated. Therefore, we have 
%----------
\vspace{0.2em}
\begin{align}~\label{eq:received_signal_delay_sensing2}
 y_{s}[n] \!=\! \!\sum\nolimits_{l=1}^{L_s}\!\mathbf{h}_{s,l}^H\mathbf{f}_{s}s_d[n \!- \!n_{s,l}]\! +\! z_s[n] 
 \!=\!\mathbf{s}_d^H[n]\mathbf{h}_s \!+\! z_s[n].   
\end{align}
%----------------

To mitigate multipath interference in delay sensing, we define $\mathbf{H}_{s,1} = [\mathbf{h}_{s,2},\mathbf{h}_{s,3},\dots,\mathbf{h}_{s,L_s}]$ and set $\mathbf{Q}_s = \mathbf{I} - \mathbf{H}_{s,1}(\mathbf{H}_{s,1}^H\mathbf{H}_{s,1})^{-1}\mathbf{H}_{s,1}^H$. Therefore, we have $\mathbf{f}_s = \mathbf{Q}_s\mathbf{b}_s$. The sensing signal in~\eqref{eq:received_signal_delay_sensing2} is simplified as
%---------
\vspace{0.2em}
\begin{align}
    y_{s}[n] = \mathbf{s}^H_d[n]\mathbf{p}(\tau_{s,1})\beta_{s,1}^*\mathbf{a}^H(\varphi_{s,1})\mathbf{f}_{s} + z_s[n]. 
\end{align}
%-----------------

\subsubsection{Maximum Likelihood}
We collect $M_2$ signals as $\mathbf{y}_{s} = [y_{s}[n],\dots,y_{s}[n+M_2-1]]^T = \mathbf{S}_d^H\mathbf{p}(\tau_{s,1})\beta\mathbf{a}^H(\varphi_{s,1})\mathbf{f}_{s} + \mathbf{z}_s$, where $\mathbf{y}_{s}\in\mathbb{C}^{M_2\times 1}$, $\mathbf{S}_d = [\mathbf{s}_d[n],\dots,\mathbf{s}_d[n+M_2-1]]\in\mathbb{C}^{M\times M_2}$, and $\mathbf{z}_s[n] = [z_s[n];\dots;z_s[n+M_2-1]]\in\mathbb{C}^{M_2\times 1}$. 

Then, the maximum likelihood method is used to estimate the amplitude and delay as
%------------
\vspace{0.2em}
\begin{align}
    \widehat{\beta},\widehat{\tau}_{s,1} = \arg\min\|\mathbf{y}_{s} - \mathbf{S}_d^H\mathbf{p}(\tau_{s,1})\beta\mathbf{a}^H(\varphi_{s,1})\mathbf{f}_{s}\|^2_2.
\end{align}
%------------

Note that for given $\tau_{s,1}$, the minimizer of the above objective with respect to $\beta$ is given by $\widehat{\beta} = \frac{\boldsymbol{\xi}^H\mathbf{y}_{s}}{\|\boldsymbol{\xi}\|^2_2}$, where $\boldsymbol{\xi} = \mathbf{S}_d^H\mathbf{p}(\tau_{s,1})\mathbf{a}^H(\varphi_{s,1})\mathbf{f}_{s}\in\mathbb{C}^{M_2\times 1}$. Substituting the amplitude into the objective, we have $\widehat{\tau}_{s,1} = \arg\max \frac{|\boldsymbol{\xi}^H\mathbf{y}_{s}|^2}{\|\boldsymbol{\xi}\|^2_2}$. The delay parameter $\tau$ is iteratively updated along the gradient-ascent
direction of $J(\tau) = \frac{|\boldsymbol{\xi}^H\mathbf{y}_s|^2}{\|\boldsymbol{\xi}\|_2^2}$,
and the estimated channel gain $\widehat{\beta}$ is finally obtained
by substituting $\widehat{\tau}_{s,1}$ into~$\widehat{\beta} = \frac{\boldsymbol{\xi}^H\mathbf{y}_{s}}{\|\boldsymbol{\xi}\|^2_2}$.

\subsubsection{CRB for Delay Estimation}
To evaluate the performance of delay estimation, we define the complex amplitude parameter as $\bm{\beta} = [\Re\{\beta\}, \Im\{\beta\}]^{T}\in\mathbb{R}^{2\times 1}$.
The vector of unknown parameters is then expressed as $\bm{\xi} = [\tau_{s,1}, \bm{\beta}^T]^T\in\mathbb{R}^{3\times 1}$. The Fisher information matrix $\mathbf{J} \in \mathbb{R}^{3 \times 3}$ corresponding to the parameter vector $\bm{\xi}$ can be computed by following \cite{He2026}. 
\begin{proposition}
    The CRB for the delay parameter $\tau_{s,1}$, $ \text{CRB}(\tau_{s,1}) = [\mathbf{J}^{-1}]_{11}$ is expressed as
    %--------------------
    \vspace{0.1em}
\begin{align}\label{CRB1}
    \text{CRB}(\tau_{s,1}) = 
    \frac{1}{|\beta|^2\zeta}\bigg(\widetilde{\mathbf{p}}^H_{\tau_{s,1}}\mathbf{R}_J\widetilde{\mathbf{p}}_{\tau_{s,1}} -\frac{|\widetilde{\mathbf{p}}^H_{\tau_{s,1}}\mathbf{R}_J\mathbf{p}(\tau_{s,1})|^2}{\mathbf{p}^H(\tau_{s,1})\mathbf{R}_J\mathbf{p}(\tau_{s,1})}\bigg)^{-1},
\end{align}
%--------------
where $\zeta\! =\! \frac{2}{\sigma_a^2}|\mathbf{a}^{H}(\varphi_{s,1})\mathbf{f}_{s}|^2$, $\mathbf{R}_J \triangleq \mathbf{S}_d\mathbf{S}_d^{H}\!\in\!\mathbb{C}^{M\times M}$, and  $\widetilde{\mathbf{p}}_{\tau_{s,1}} \!= \!\frac{\partial \mathbf{p}(\tau_{s,1})}{\partial \tau_{s,1}} \!=\! -B\mathbf{d}(\tau_{s,1})\in\mathbb{C}^{M\times 1}$, while $[\mathbf{d}(\tau_{s,1})]_k\! = \frac{d\, \operatorname{sinc}(k + B(\eta_s - \tau_{s,1}))}{d \tau_{s,1}}
$ with $\operatorname{sinc}(x) = \frac{\sin(\pi x)}{\pi x}$.
\end{proposition}

\begin{Proof}
The proof follows by deriving $\mathbf{J}$, which is omitted due to space limitations.
\end{Proof}

\section{Downlink Secure Communicaitons}~\label{sec:secure_communication}
In the second stage, the BS transmits confidential data to multiple UEs while continuing to estimate the target’s delay. Due to multipath propagation, DAM is employed to time-align desired signal components at each UE, while path-based precoding is used to direct energy and mitigate interference. This ensures that signals arrive aligned with the strongest path, enhancing reliability and signal strength. Although Eve attempts to intercept all UEs, the user-specific delay alignment introduces timing mismatches at Eve, weakening its received signal and giving legitimate users a performance advantage.

\subsection{Received Signal Model at UE $k$}
The received signal at the $k$-th UE is given by 
\begin{align}~\label{eq:received_signal1}
    y_{c,k}[n] = \mathbf{h}_{c,k}^H[n]\odot\mathbf{x}_c[n] + \mathbf{h}_{c,k}^H[n]\odot\mathbf{x}_s[n] + z_k[n],
\end{align}
where $z_k[n]$ is the noise with variance $\sigma_k^2$.
\begin{figure*}
    \begin{align}~\label{eq:received_signal2}
    y_{c,k}[n]  &= \underbrace{\sum\nolimits_{l=1}^{L_{c,k}}\mathbf{h}_{c,kl}^H\mathbf{f}_{c,kl}s_{c,k}[n-n_{c,k}^{\star}]}_{\text{Disired signal}} + \underbrace{\sum\nolimits_{l=1}^{L_{c,k}}\sum\nolimits_{l^{\prime}\neq l}^{L_{c,k}}\mathbf{h}_{c,kl}^H\mathbf{f}_{c,kl^{\prime}}s_{c,k}[n - n_{c,k}^{\star} - \Delta_{kl,kl^{\prime}}]}_{\text{ISI}} \notag\\
               &+ \underbrace{\sum\nolimits_{l=1}^{L_{c,k}}\sum\nolimits_{k^{\prime}\neq k}^{K}\sum\nolimits_{l^{\prime}=1}^{L_{c,k^{\prime}}}\mathbf{h}_{c,kl}^H\mathbf{f}_{c,k^{\prime}l^{\prime}}s_{c,k^{\prime}}[n - n_{c,k^{\prime}}^{\star} - \Delta_{kl,k^{\prime}l^{\prime}}]}_{\text{IUI}} + \underbrace{\sum\nolimits_{l=1}^{L_{c,k}}\mathbf{h}_{c,kl}^H\mathbf{f}_{s}s_d[n-n_{c,kl}]}_{\text{Sensing interference}} + z_k[n]  
\end{align}
\hrulefill
\vspace{-1.2em}
\end{figure*}
We apply delay pre-compensation as $\kappa_{c,k^{\prime} l^{\prime}} = n_{c,k^{\prime}}^{\star} - n_{c,k^{\prime} l^{\prime}}$, where $n_{c,k^{\prime}}^{\star}$ denotes the synchronization reference time for the $k^{\prime}$-th UE, e.g., the arrival time of its strongest path. By applying the compensation $\kappa_{c,k^{\prime} l^{\prime}}$ to each path $l^{\prime}$, all multipath components are time-aligned and coherently combined at the receiver. The received signal in~\eqref{eq:received_signal1} is extended in~\eqref{eq:received_signal2} at the top of next page, where $\Delta_{kl,k^{\prime}l^{\prime}} = n_{c,kl} - n_{c,k^{\prime}l^{\prime}}$ is the delay difference between path $l$ of the $k$-th UE and path $l^{\prime}$ of the $k^{\prime}$-th UE. For a given user pair $(k,k^{\prime})$, the differences satisfy $\Delta_{kl,k^{\prime}l^{\prime}}\in[\Delta_{kk^{\prime},\text{min}},\Delta_{kk^{\prime},\text{max}}]$, where $\Delta_{kk^{\prime},\text{max}} = n_{c,k,\text{max}} - n_{c,k^{\prime},\text{min}}$, $\Delta_{kk^{\prime},\text{min}} = n_{c,k,\text{min}} - n_{c,k^{\prime},\text{max}}$, with $n_{c,k,\min}$ and $n_{c,k,\max}$ denoting the earliest and latest delays, respectively.

To derive the SINR of~\eqref{eq:received_signal2} with residual ISI, IUI and sensing interference, the equation needs to be reformulated by grouping those interfering symbols with identical delay differences since they correspond to identical symbols. 
We define
\begin{align}
    \mathbf{g}_{k,k^{\prime}l^{\prime}}[i] = 
    \left\{\begin{matrix}
        \mathbf{h}_{c,kl}, & \text{if}~\exists~l\in[1, L_{c,k}],~\text{s.t.}~\Delta_{kl,k^{\prime}l^{\prime}} = i,              \notag\\
        \mathbf{0},   &\text{otherwise},
    \end{matrix} \right.
\end{align}
where $l^{\prime}\in[1, L_{c,k^{\prime}}]$ and $\mathbf{g}_{k,k^{\prime}l^{\prime}}[i]\in\mathbb{C}^{N\times 1}$. Thus, \eqref{eq:received_signal2} can be further expanded as in \eqref{eq:received_signal3}, shown on top of the next page.
\begin{figure*}
    \begin{align}~\label{eq:received_signal3}
    y_{c,k}[n] &= \sum\nolimits_{l=1}^{L_{c,k}}\mathbf{h}_{c,kl}^H\mathbf{f}_{c,kl}s_{c,k}[n-n_{c,k}^{\star}] + \sum\nolimits_{i=\Delta_{kk,\text{min}},i\neq 0}^{\Delta_{kk,\text{max}}}\sum\nolimits_{l^{\prime}=1}^{L_{c,k}}\mathbf{g}_{k,kl^{\prime}}^H[i]\mathbf{f}_{c,kl^{\prime}}s_{c,k}[n - n_{c,k}^{\star} - i] \notag\\
               &+ \sum\nolimits_{k^{\prime}\neq k}^K\sum\nolimits_{i=\Delta_{kk^{\prime},\text{min}}}^{\Delta_{kk^{\prime},\text{max}}}\sum\nolimits_{l^{\prime}=1}^{L_{c,k^{\prime}}}\mathbf{g}^H_{k,k^{\prime}l^{\prime}}[i]\mathbf{f}_{c,k^{\prime}l^{\prime}}s_{c,k^{\prime}}[n - n_{c,k^{\prime}}^{\star} - i] + \sum\nolimits_{l=1}^{L_{c,k}}\mathbf{h}_{c,kl}^H\mathbf{f}_{s}s_d[n-n_{c,kl}] + z_k[n]
\end{align}
\hrulefill
\vspace{-1.2em}
\end{figure*}

Based on~\eqref{eq:received_signal3}, the SINR at UE $k$ is given by
\begin{align}~\label{eq:SINR_UE}
    \gamma_k \!=\! \frac{\bar{\mathbf{f}}_{c,k}^H\mathbf{R}_{c,k}\bar{\mathbf{f}}_{c,k}}{\bar{\mathbf{f}}_{c,k}^H\mathbf{A}_{k,k}\bar{\mathbf{f}}_{c,k} \!\!+\! \sum\nolimits_{k^{\prime}\neq k}^K\bar{\mathbf{f}}_{c,k^{\prime}}^H\mathbf{B}_{k,k^{\prime}}\bar{\mathbf{f}}_{c,k^{\prime}} \!\!+\! \mathbf{f}_{s}^H\mathbf{B}_{s,k}\mathbf{f}_{s} \!\!+\! 1},
\end{align}
%---------------
where $\bar{\mathbf{f}}_{c,k} \!\!=\!\! [\mathbf{f}_{c,k1}^T,\dots,\mathbf{f}_{c,kL_{c,k}}^T]^T\!\!\in\!\!\mathbb{C}^{L_{c,k}N\times 1}$; $\mathbf{R}_{c,k} \!=\!\! \bar{\mathbf{h}}_{c,k}\bar{\mathbf{h}}_{c,k}^H/\sigma_k^2\in\mathbb{C}^{L_{c,k}N\times L_{c,k}N}$ with $\bar{\mathbf{h}}_{c,k} \!\!= \!\![\mathbf{h}_{c,k1}^T,\dots,\mathbf{h}_{c,kL_{ck}}^T]^T\!\in\!\mathbb{C}^{L_{c,k}N\times 1}$; $\mathbf{A}_{k,k} = \bar{\mathbf{G}}_{k,k}\bar{\mathbf{G}}_{k,k}^H/\sigma_k^2$ with $\bar{\mathbf{G}}_{k,k} = [\bar{\mathbf{g}}_{k,k}[\Delta_{kk,\text{min}}],\dots,\bar{\mathbf{g}}_{k,k}[\Delta_{kk,\text{max}}]]\in\mathbb{C}^{L_{c,k}N,\Delta_{kk,\textrm{span}}}$, $\bar{\mathbf{g}}_{k,k}[i] = \big[\mathbf{g}^T_{k,k1}[i],\dots,\mathbf{g}^T_{k,kL_{c,k}}[i]\big]^T\in\mathbb{C}^{L_{c,k}N\times 1}$ and $\Delta_{kk,\textrm{span}} = \Delta_{kk,\text{max}} - \Delta_{kk,\text{min}}$; $\mathbf{B}_{k,k^{\prime}} = \widetilde{\mathbf{G}}_{k,k^{\prime}}\widetilde{\mathbf{G}}_{k,k^{\prime}}^H/\sigma_k^2$ with $\widetilde{\mathbf{G}}_{k,k^{\prime}} = [\widetilde{\mathbf{g}}_{k,k^{\prime}}[\Delta_{kk^{\prime},\text{min}}],\dots,\widetilde{\mathbf{g}}_{k,k^{\prime}}[\Delta_{kk^{\prime},\text{max}}]]\in\mathbb{C}^{L_{c,k^{\prime}}N,\Delta_{kk^{\prime},\textrm{span}}}$, $\widetilde{\mathbf{g}}_{k,k^{\prime}}[i]\!\! =\!\! \big[\mathbf{g}^T_{k,k^{\prime}1}[i],\dots,\mathbf{g}^T_{k,k^{\prime}L_{c,k^{\prime}}}[i]\big]^T\!\!\in\!\!\mathbb{C}^{L_{c,k^{\prime}}N\times 1}$, and $\Delta_{kk^{\prime},\textrm{span}} = \Delta_{kk^{\prime},\text{max}} - \Delta_{kk^{\prime},\text{min}} + 1$ for $k^{\prime}\neq k$, while $\mathbf{B}_{s,k} = \mathbf{H}_{c,k}\mathbf{H}_{c,k}^H/\sigma_k^2$ with $\mathbf{H}_{c,k} = [\mathbf{h}_{c,k1},\dots,\mathbf{h}_{c,kL_k}]\in\mathbb{C}^{N\times L_{c,k}}$.

\subsection{Received Signal Model at Eve}
The received signal at Eve is given by 
\begin{align}~\label{eq:received_signal_Eve1}
    y_{e,k}[n] = \mathbf{h}_{e}^H[n]\odot\mathbf{x}_s[n] + \mathbf{h}_{e}^H[n]\odot\mathbf{x}_c[n] + z_e[n],
\end{align}
where $z_e[n]$ is the noise with variance $\sigma_e^2$.
Since $\kappa_{c,k^{\prime}l^{\prime}} = n_{c,k^{\prime}}^{\star} - n_{c,k^{\prime}l^{\prime}}$, the received signal in~\eqref{eq:received_signal_Eve1} is expanded as
%------------
    \begin{align}~\label{eq:received_signal_Eve2}
    y_{e,k}[n] &=\sum\limits_{l=1}^{L_e}\sum\limits_{k^{\prime}=1}^K\sum\limits_{l^{\prime}=1}^{L_c,k^{\prime}}\mathbf{h}_{e,l}^H\mathbf{f}_{c,k^{\prime}l^{\prime}}s_{c,k^{\prime}}[n-n_{c,k^{\prime}}^{\star} - \Delta_{el,k^{\prime}l^{\prime}}]  
    \nonumber\\
    &+\sum\nolimits_{l=1}^{L_e}\mathbf{h}_{e,l}^H\mathbf{f}_{s}s_d[n - n_{e,l}] + z_e[n],
\end{align}
%------
where $\Delta_{el,k^{\prime}l^{\prime}} = n_{e,l} - n_{c,k^{\prime}l^{\prime}}$ is the delay difference between the $l$-th path of Eve to the $l^{\prime}$-th path of the $k$-th UE.  We have $\Delta_{el,k^{\prime}l^{\prime}}\in[\Delta_{ek^{\prime},\text{min}},\Delta_{ek^{\prime},\text{max}}]$, where $\Delta_{ek^{\prime},\text{min}} = n_{e,\text{min}} - n_{c,k^{\prime},\text{max}}$ and $\Delta_{ek^{\prime},\text{max}} = n_{e,\text{max}} - n_{c,k^{\prime},\text{min}}$, with $n_{e,\min}$ and $n_{e,\max}$ denoting Eve's earliest and latest delays, respectively.

It is worth noting that multiple replicas of the same transmitted symbol in Eve’s received signal may share an identical delay difference. In this case, Eve can synchronize to that delay difference and potentially intercept the confidential message. 
Therefore, these symbol components should be properly grouped. 
For any delay difference $i$ within the interval $[\Delta_{ek^{\prime},\text{min}}, \Delta_{ek^{\prime},\text{max}}]$, we define the following channel
\begin{align}
    \mathbf{g}_{e,k^{\prime}l^{\prime}}[i] = \left\{\begin{matrix}
    \mathbf{h}_{e,l}, &\text{if}~\exists~l\in[1, L_e],~ \text{s.t.}~\Delta_{el,k^{\prime}l^{\prime}} = i, \notag\\
    \mathbf{0},         &\text{otherwise},
    \end{matrix} \right.
\end{align}
where $l^{\prime}\in[1, L_{c,k^{\prime}}]$.

Define $i^{\star} = \arg_{i\in[\Delta_{ek,\text{min}},\Delta_{ek,\text{max}}]} \max \sum_{l^{\prime}=1}^{L_{c,k}}\|\mathbf{g}_{e,kl^{\prime}}[i]\|_2^2$ as the time index at which Eve attains the maximum received power for the $k$-th UE. %; Eve synchronizes to $n_{c,k}^{\star}+i^{\star}$ to intercept the $k$-th UE’s transmission. 
The corresponding received signal %at Eve when eavesdropping on the $k$-th UE 
can be equivalently expressed as~\eqref{eq:received_signal_Eve3}, on top of the next page.
\begin{figure*}
    \begin{align}~\label{eq:received_signal_Eve3}
    y_{e,k}[n] &= \sum\nolimits_{l^{\prime}=1}^{L_{c,k}}\mathbf{g}_{e,kl^{\prime}}^H[i]\mathbf{f}_{c,kl^{\prime}}s_{c,k}[n-n_{c,k}^{\star} - i^{\star}] + \sum\nolimits_{i=\Delta_{ek,\text{min}},i\neq i^{\star}}^{\Delta_{ek,\text{max}}}\sum\nolimits_{l^{\prime}=1}^{L_{c,k}}\mathbf{g}_{e,kl^{\prime}}^H[i]\mathbf{f}_{c,kl^{\prime}}s_{c,k}[n-n_{c,k}^{\star} - i] \notag\\           
               &+ \sum\nolimits_{i=\Delta_{ek^{\prime},\text{min}}}^{\Delta_{ek^{\prime},\text{max}}}\sum\nolimits_{k^{\prime}\neq k}^K\sum\nolimits_{l^{\prime}=1}^{L_{c,k^{\prime}}}\mathbf{g}_{e,k^{\prime}l^{\prime}}^H[i]\mathbf{f}_{c,k^{\prime}l^{\prime}}s_c[n-n_{c,k^{\prime}}^{\star} - i] + \sum\nolimits_{l=1}^{L_e}\mathbf{h}_{e,l}^H\mathbf{f}_{s}s[n - n_{e,l}] + z_e[n] 
\end{align}
\hrulefill
\vspace{-1.5em}
\end{figure*}
%-------------

Based on~\eqref{eq:received_signal_Eve3}, the SINR at Eve can be written as
\begin{align}
    \gamma_{e,k} \!=\! \frac{\bar{\mathbf{f}}_{c,k}^H\mathbf{R}_{e,k}\bar{\mathbf{f}}_{c,k}}{\bar{\mathbf{f}}_{c,k}^H\mathbf{A}_{e,k}\bar{\mathbf{f}}_{c,k} \!\!+\! \sum_{k^{\prime}\neq k}^K\bar{\mathbf{f}}_{c,k^{\prime}}^H\mathbf{B}_{e,k^{\prime}}\bar{\mathbf{f}}_{c,k^{\prime}} \!\!+\! \mathbf{f}_{s}^H\mathbf{B}_{es}\mathbf{f}_s \!\!+ \!1},
\end{align}
%-------------------
where $\mathbf{R}_{e,k} = \bar{\mathbf{g}}_{e,k}[i^{\star}]\bar{\mathbf{g}}_{e,k}^H[i^{\star}]/\sigma_e^2\in\mathbb{C}^{L_{c,k}N\times L_{c,k}N}$ with $\bar{\mathbf{g}}_{e,k}[i] = \big[\mathbf{g}^T_{e,k1}[i],\dots,\mathbf{g}^T_{e,kL_{c,k}}[i]\big]^T\in\mathbb{C}^{L_{c,k}N\times 1}$; 
$\mathbf{A}_{e,k} = \bar{\mathbf{G}}_{e,k}\bar{\mathbf{G}}^H_{e,k}/\sigma_e^2$ with $\bar{\mathbf{G}}_{e,k} = \big[\bar{\mathbf{g}}_{e,k}[\Delta_{ek,\text{min}}],\dots,\bar{\mathbf{g}}_{e,k}[\Delta_{ek,\text{max}}]\big]\in\mathbb{C}^{L_{c,k}N,\Delta_{ek,\textrm{span}}}$ and
$\Delta_{ek,\textrm{span}} = \Delta_{ek,\text{max}} - \Delta_{ek,\text{min}}$; 
$\mathbf{B}_{e,k^{\prime}} = \widetilde{\mathbf{G}}_{e,k^{\prime}}\widetilde{\mathbf{G}}^H_{e,k^{\prime}}/\sigma_e^2$ with $\widetilde{\mathbf{G}}_{e,k^{\prime}} = \big[\widetilde{\mathbf{g}}_{e,k^{\prime}}[\Delta_{ek^{\prime},\text{min}}],\dots,\widetilde{\mathbf{g}}_{e,k^{\prime}}[\Delta_{ek^{\prime},\text{max}}]\big]\in\mathbb{C}^{L_{c,k^{\prime}}N,\Delta_{ek^{\prime},\textrm{span}}}$, $\widetilde{\mathbf{g}}_{e,k^{\prime}}[i] = \big[\mathbf{g}^T_{e,k^{\prime}1}[i],\dots,\mathbf{g}^T_{e,k^{\prime}L_{c,k^{\prime}}}[i]\big]^T\in\mathbb{C}^{L_{c,k^{\prime}}N\times 1}$, and $\Delta_{ek^{\prime},\text{\textrm{span}}} = \Delta_{ek^{\prime},\text{max}} - \Delta_{ek^{\prime},\text{min}}$ + 1 for $k^{\prime} \neq k$, while
$\mathbf{B}_{es} = \mathbf{H}_{e}\mathbf{H}_{e}^H/\sigma_e^2$ with $\mathbf{H}_{e} = [\mathbf{h}_{e,1},\dots,\mathbf{h}_{e,L_e}]\in\mathbb{C}^{N\times L_e}$. 

\subsection{Secrecy Spectral Efficiency}
The SSE quantifies how much confidential information can be transmitted reliably to a legitimate user, while remaining information-theoretically secure from an Eve; %it is the difference between the achievable rates of the legitimate link and the eavesdropper. Accordingly, the secrecy rate of the $k$-th UE is
\begin{align}~\label{eq:SR_k}
 R_k &= \frac{n_c -2n_{B,\text{max}}}{n_c}\Big[\log_2(1 + \gamma_{k}) - \log_2(1 + \gamma_{e,k})\Big]^{+},
\end{align}
where $n_c \!=\! T_c/T_s$ is the number of symbol intervals within one channel coherence block $T_c$. The guard interval length is modeled as $\bar{n}_{B,\text{max}} = n_{B,\text{max}} + (n_{B,\text{max}} - n_{B,\text{min}})\approx 2n_{B,\text{max}}$, leading to the factor $(n_c-2n_{B,\max})/n_c$.

%Here, $\gamma_{k}$ and $\gamma_{e,k}$ denote the received SINR at the $k$-th UE and at the Eve, respectively. As indicated by \eqref{eq:SR_k}, $R_k$ depends on the transmit precoders $\mathbf{f}_{c,kl}$ and the sensing precoder $\mathbf{f}_s$. In the next subsection, we design $\{\mathbf{f}_{c,kl}\}$ and $\mathbf{f}_s$ to optimize SR under the DAM-enabled ISAC constraints.
%------------------------------
\section{Path-Based Precoding for Interference-Free Transmission}~\label{sec: ZF_Optimized}
%------------------------------
Building on the SSE expression in~\eqref{eq:SR_k}, we design transmit precoders for the DAM-enabled ISAC system. Since the received SINR is degraded by sensing interference, ISI, and IUI, we adopt a path-based ZF strategy: per-path precoders suppress ISI and IUI, while DAM aligns the desired signal in time. The objective is to maximize the worst-case user SSE, subject to a delay estimation accuracy constraint expressed via the CRB. The resulting max–min SSE optimization problem is given by
% Building on the SR in~\eqref{eq:SR_k}, we now design the transmit precoders under the DAM-enabled ISAC setting. Since the received SINR is degraded by sensing interference, ISI, and IUI, we adopt a path-based ZF strategy: per-path precoders are constructed to suppress ISI/IUI while DAM aligns the desired signal in time. The design goal is to maximize the worst-user secrecy rate subject to a delay-estimation accuracy constraint expressed via the CRB. The max–min SR problem is given by
\begin{subequations}
\begin{align}
    \text{(P1)}~\mathop{\max_{\bar{\mathbf{f}}_{c,k},\mathbf{f}_s}~\min_k}&~\{R_{k}\} \\
\text{s.t.}&~\text{CRB}(\tau_{s,1}) \leq \Gamma,  \label{constraint:CRB}\\ 
           &~\sum\nolimits_k\|\bar{\mathbf{f}}_{c,k}\|^2_2 + \|\mathbf{f}_s\|^2_2 \leq P, \label{constraint:power} \\
           &~\mathbf{f}_s = \bar{\mathbf{Q}}_s\bar{\mathbf{b}}_s,~\bar{\mathbf{f}}_{c,k} = \bar{\mathbf{Q}}_{c,k}\bar{\mathbf{b}}_{c,k}. \label{constraint:sensing_inter_mitigation}% \\
          % &~\bar{\mathbf{f}}_{c,k} = \bar{\mathbf{Q}}_{c,k}\bar{\mathbf{b}}_{c,k} \label{constraint:comm_inter_mitigation}.
\end{align}    
\end{subequations}
Here, $\Gamma$ is the CRB threshold for delay estimation, and $P$ is the total transmit power. The constraints above ensure accurate sensing, power compliance, and mitigation of multipath and communication-induced interference. To simplify (P1), we adopt ZF-based block-diagonal precoders $\bar{\mathbf{Q}}_{c,k}$ and $\bar{\mathbf{Q}}_s$ to null ISI, IUI, and sensing interference. With this approach, $\bar{\mathbf{b}}_s$ and $\bar{\mathbf{b}}_{c,k}$ are optimized to maximize SSE under CRB and power constraints.

In order to mitigate the interference from the communications signal to the sensing signal, we design ZF precoding. We should have $\mathbf{h}_{s,l^{\prime}}^H\mathbf{f}_{c,kl} = 0,~\text{for}~l^{\prime}=1,\dots,L_s$. Also, the precoding vector should mitigate the ISI and IU interference. We have the requirements as follows: $\mathbf{h}_{c,kl^{\prime}}^H\mathbf{f}_{c,kl} = 0,~\text{for}~l^{\prime}\neq l$ and $\mathbf{h}_{c,k^{\prime},l^{\prime}}^H\mathbf{f}_{c,kl} = \mathbf{0},~\text{for}~k\neq k^{\prime}$.

Therefore, the ZF precoding vector is given by
%------------
\begin{align}
    \mathbf{f}_{c,kl} = \mathbf{Q}_{c,kl}\mathbf{b}_{c,kl},
\end{align}
%--------------
where $\mathbf{Q}_{c,kl} = \mathbf{I}_N - \mathbf{H}_{sc,kl}(\mathbf{H}_{sc,kl}^H\mathbf{H}_{sc,kl})^{-1}\mathbf{H}_{sc,kl}^H$, with $\mathbf{H}_{sc,kl} = [\mathbf{h}_{s,1}, \dots, \mathbf{h}_{s,L_s}, \dots,\mathbf{h}_{c,k(l-1)}, \mathbf{h}_{c,k(l+1)}\dots, \mathbf{h}_{c,KL_K}]$. %We define $\bar{\mathbf{f}}_{c,k} = [\mathbf{f}_{c,k1}^T,\dots,\mathbf{f}_{c,kL_{c,k}}^T]^T\in\mathbb{C}^{L_{c,k}N\times 1}$. 
Thus, $\bar{\mathbf{f}}_{c,k} = \bar{\mathbf{Q}}_{c,k}\bar{\mathbf{b}}_{c,k}$, where $\bar{\mathbf{b}}_{c,k} = [\mathbf{b}_{c,k1}^T,\dots,\mathbf{b}_{c,kL_{c,k}}^T]^T\in\mathbb{C}^{L_{c,k}N\times 1}$ and $\bar{\mathbf{Q}}_{c,k} = \operatorname{blkdiag}(\mathbf{Q}_{c,k1},\dots,\mathbf{Q}_{c,kL_{c,k}})\in\mathbb{C}^{L_{c,k}N\times L_{c,k}N}$.
To mitigate sensing interference for secure communications, we should have $\mathbf{h}_{c,kl}^H\mathbf{f}_{s} = 0$. We define $\widetilde{\mathbf{H}}_{s,1} = [\mathbf{h}_{s,2},\mathbf{h}_{s,3},\dots,\mathbf{h}_{s,L_s},\mathbf{h}_{c,1,1},\dots,\mathbf{h}_{c,k,l},\dots,\mathbf{h}_{c,K,L_{c,K}}]$ and set $\bar{\mathbf{Q}}_s = \mathbf{I} - \widetilde{\mathbf{H}}_{s,1}(\widetilde{\mathbf{H}}_{s,1}^H\widetilde{\mathbf{H}}_{s,1})^{-1}\widetilde{\mathbf{H}}_{s,1}^H$. Therefore, we have $\mathbf{f}_s = \bar{\mathbf{Q}}_s\bar{\mathbf{b}}_s$. Accordingly, the signal-to-noise ratio (SNR) of the $k$-th UE is given by
%-----------
\begin{align}~\label{eq:gamk}
   \gamma_k = \bar{\mathbf{b}}_{c,k}^H\bar{\mathbf{R}}_{c,k}\bar{\mathbf{b}}_{c,k}, 
\end{align}
where $\bar{\mathbf{R}}_{c,k} = \bar{\mathbf{Q}}_{c,k}^H\mathbf{R}_{c,k}\bar{\mathbf{Q}}_{c,k}\in\mathbb{C}^{L_{c,k}N\times L_{c,k}N}$.
%--------------------

Moreover, the SINR of the Eve is given by $\gamma_{e,k} =$
%-------------
\vspace{-0.1em}
\begin{align}~\label{eq:game}
   % \gamma_{e,k} \!=\!
\frac{\bar{\mathbf{b}}_{c,k}^H\bar{\mathbf{R}}_{e,k}\bar{\mathbf{b}}_{c,k}}{\bar{\mathbf{b}}_{c,k}^H\bar{\mathbf{A}}_{e,k}\bar{\mathbf{b}}_{c,k} \!+ \sum\nolimits_{k^{\prime}\neq k}^K\bar{\mathbf{b}}_{c,k^{\prime}}^H\bar{\mathbf{B}}_{e,k^{\prime}}\bar{\mathbf{b}}_{c,k^{\prime}} \!+\! \bar{\mathbf{b}}_{s}^H\bar{\mathbf{B}}_{es}\bar{\mathbf{b}}_s \!+\!\! 1}, 
\end{align}
where $\bar{\mathbf{R}}_{e,k} = \bar{\mathbf{Q}}_{c,k}^H\mathbf{R}_{e,k}\bar{\mathbf{Q}}_{c,k}\!\in\!\mathbb{C}^{L_{c,k}N\times L_{c,k}N}$, $\bar{\mathbf{A}}_{e,k} \!=\! \bar{\mathbf{Q}}_{c,k}^H\mathbf{A}_{e,k}\bar{\mathbf{Q}}_{c,k}\!\in\!\mathbb{C}^{L_{c,k}N\times L_{c,k}N}$, $\bar{\mathbf{B}}_{e,k^{\prime}} = \bar{\mathbf{Q}}_{c,k^{\prime}}^H\mathbf{B}_{e,k^{\prime}}\bar{\mathbf{Q}}_{c,k^{\prime}}\!\in\!\mathbb{C}^{L_{c,k^{\prime}}N\times L_{c,k^{\prime}}N}$ and $\bar{\mathbf{B}}_{es}\! =\! \bar{\mathbf{Q}}_{s}\mathbf{B}_{es}\bar{\mathbf{Q}}_{s}\!\in\!\mathbb{C}^{N\times N}$.
%---------------

%We have the SR of the $k$-th UE as
%\begin{figure*}
%    \begin{align}
%    C_{b,k} - C_{e,k} = \log_2\Big(1 + \bar{\mathbf{f}}_{c,k}^H\bar{\mathbf{R}}_{c,k}\bar{\mathbf{f}}_{c,k}\Big) - \log_2\Big(1 + \frac{\bar{\mathbf{b}}_{c,k}^H\bar{\mathbf{R}}_{e,k}\bar{\mathbf{b}}_{c,k}}{\bar{\mathbf{b}}_{c,k}^H\bar{\mathbf{A}}_{e,k}\bar{\mathbf{b}}_{c,k} + \sum\nolimits_{k_2\neq k}\bar{\mathbf{b}}_{c,k_2}^H\bar{\mathbf{B}}_{e,k_2}\bar{\mathbf{b}}_{c,k_2} + \mathbf{b}_{s}^H\bar{\mathbf{B}}_{es}\mathbf{b}_{s} + 1}\Big)
%\end{align}
%\hrulefill
%\end{figure*}
%------------------------
% \subsection{Optimization problem}~\label{subsec: Precoding_ZF_Optimized}
%--------------

Now, for given SINR expressions under ZF-based precoding design, we look into sensing precoding design problem.  Let  $\Gamma_2 \triangleq \frac{1}{2|\beta|^2\Gamma G}$ and $G = \widetilde{\mathbf{p}}^H_{\tau_{s,1}}\mathbf{R}_J\widetilde{\mathbf{p}}_{\tau_{s,1}} -\frac{|\widetilde{\mathbf{p}}^H_{\tau_{s,1}}\mathbf{R}_J\mathbf{p}(\tau_{s,1})|^2}{\mathbf{p}^H(\tau_{s,1})\mathbf{R}_J\mathbf{p}(\tau_{s,1})}$. Then, the constraint~\eqref{constraint:CRB} can be expressed as
%-------------
\begin{align}~\label{constraint:CRB2}
\bar{\mathbf{b}}_{s}^H\bar{\mathbf{Q}}_s^H\mathbf{A}_{s,1}\bar{\mathbf{Q}}_s\bar{\mathbf{b}}_s\geq \Gamma_2\sigma_a^2,
\end{align}
%-----------
which is a non-convex constraint. Here, $\mathbf{A}_{s,1} = \mathbf{a}_{s,1}\mathbf{a}_{s,1}^H$. To address this issue, we apply successive convex approximation (SCA). For clarity, we denote by a superscript ($n$) the value of a variable obtained after ($n-1$) iterations, where $n \geq 1$. Specifically to handle~\eqref{constraint:CRB2}, we use the concave lower bound of the right-hand-side to get 
%-----------
\begin{align}~\label{eq:cct1}
    &2\Re\big\{\!\bar{\mathbf{b}}_{s}^{\!H}\bar{\mathbf{Q}}_s^{\!H}\mathbf{A}_{s,1}\bar{\mathbf{Q}}_s\bar{\mathbf{b}}_{s}^{(n)}\!\big\}\!-\!\!\bar{\mathbf{b}}_{s}^{(n) H}\bar{\mathbf{Q}}_s^{\!H}\mathbf{A}_{s,1}\bar{\mathbf{Q}}_s\bar{\mathbf{b}}_{s}^{(n)} \!\!\geq\! \Gamma_2\sigma_a^2.
\end{align}
%---------------------------

Now, by invoking~\eqref{eq:SR_k},~\eqref{eq:gamk}, and~\eqref{eq:game}, and then introducing the slack variables, $\gamma$, $\qt_k=\{t_{1,k}, t_{2,k}, t_{3,k}\}$ and $\qq_k=\{q_{1,k}, q_{2,k}, q_{3,k}\}$, the optimization problem $\text{(P2)}$ is recast as
%---------------
\begin{subequations}
\begin{align}
    \text{(P2)}~&\mathop{\max_{\bar{\mathbf{b}}_{c,k}, \bar{\mathbf{b}}_s, \mathbf{t}_k, \mathbf{q}_k}}~\gamma \\
\text{s.t.}~&~t_{1,k} - t_{2,k} + t_{3,k} \geq \gamma, \label{const0}\\
            %---------------
             &~\bar{\mathbf{b}}_{c,k}^H\bar{\mathbf{R}}_{c,k}\bar{\mathbf{b}}_{c,k} + 1 \geq q_{1,k}, \label{const2}\\
             %---- 
             &~\bar{\mathbf{b}}_{c,k}^H(\bar{\mathbf{A}}_{e,k} + \bar{\mathbf{R}}_{e,k})\bar{\mathbf{b}}_{c,k} + \bar{\mathbf{b}}_{s}^H\bar{\mathbf{B}}_{es}\bar{\mathbf{b}}_{s} \notag\\
            &~+ \sum\nolimits_{k^{\prime}\neq k}\bar{\mathbf{b}}_{c,k^{\prime}}^H\bar{\mathbf{B}}_{e,k^{\prime}}\bar{\mathbf{b}}_{c,k^{\prime}} + 1 \leq q_{2,k}, \label{const6}\\            
            %--------
            &~\bar{\mathbf{b}}_{c,k}^H\bar{\mathbf{A}}_{e,k}\bar{\mathbf{b}}_{c,k} + \sum\nolimits_{k^{\prime}\neq k}^K\bar{\mathbf{b}}_{c,k^{\prime}}^H\bar{\mathbf{B}}_{e,k^{\prime}}\bar{\mathbf{b}}_{c,k^{\prime}} \notag\\
            %----
            &~+ \bar{\mathbf{b}}_{s}^H\bar{\mathbf{B}}_{es}\bar{\mathbf{b}}_{s} + 1 \geq q_{3,k}, \label{const8}\\
            %----
            % &~\bar{\mathbf{b}}_{s}^H\bar{\mathbf{Q}}_s^H\mathbf{a}_{s,1}\mathbf{a}_{s,1}^H\bar{\mathbf{Q}}_s\bar{\mathbf{b}}_s\geq \Gamma_2\sigma_a^2,  \label{const9}\\ 
            %--------------
            &~\sum\nolimits_k\|\bar{\mathbf{Q}}_{c,k}\bar{\mathbf{b}}_{c,k}\|^2_2 + \|\bar{\mathbf{Q}}_s\bar{\mathbf{b}}_s\|_2^2 \leq P, \label{const10}\\
             %----
             &~\log_2( q_{1,k}) \geq t_{1,k},     \label{const1}\\
             %----
            &~\log_2(q_{2,k}) \leq t_{2,k}, \label{const5}\\
            %------           
            &~\log_2(q_{3,k}) \geq t_{3,k}, \label{const7}\\
            %------------
            &~\eqref{eq:cct1}.
\end{align}    
\end{subequations}
%---------------
% where constraint~\eqref{constraint_CRB} was transferred to \eqref{const9}, with $\Gamma_2 = \frac{1}{2|\beta|^2\Gamma G}$ and $G = \widetilde{\mathbf{p}}^H_{\tau_{s,1}}\mathbf{R}_J\widetilde{\mathbf{p}}_{\tau_{s,1}} -\frac{|\widetilde{\mathbf{p}}^H_{\tau_{s,1}}\mathbf{R}_J\mathbf{p}(\tau_{s,1})|^2}{\mathbf{p}^H(\tau_{s,1})\mathbf{R}_J\mathbf{p}(\tau_{s,1})}$.
%in \eqref{const6}, \eqref{const10}, \eqref{const1},  \eqref{const7},  are convex. The constraints
Problem \text{(P2)} remains challenging due to the presence of non-convex terms in constraints \eqref{const2}, \eqref{const8}, and \eqref{const5}. To address these non-convexities, we employ the SCA technique. Consequently, constraints \eqref{const2}, \eqref{const8}, and \eqref{const5} can be reformulated into convex forms as follows:
%------------------
\begin{subequations}
    \begin{align}
        &2\Re\{(\bar{\mathbf{b}}_{c,k})^H\bar{\mathbf{R}}_{c,k}\bar{\mathbf{b}}_{c,k}^{(n)}\} - (\bar{\mathbf{b}}_{c,k}^{(n)})^H\bar{\mathbf{R}}_{c,k}\bar{\mathbf{b}}_{c,k}^{(n)}  + 1 \geq q_{1,k},\\
    %---
        &2\Re\{(\bar{\mathbf{b}}_{c,k})^H\bar{\mathbf{A}}_{e,k}\bar{\mathbf{b}}_{c,k}^{(n)}\} - (\bar{\mathbf{b}}_{c,k}^{(n)})^H\bar{\mathbf{A}}_{e,k}\bar{\mathbf{b}}_{c,k}^{(n)}  \notag\\
        &+ \sum\nolimits_{k^{\prime}\neq k}^K [2\Re\{(\bar{\mathbf{b}}_{c,k^{\prime}})^H\bar{\mathbf{B}}_{e,k^{\prime}}\bar{\mathbf{b}}_{c,k^{\prime}}^{(n)}\} - (\bar{\mathbf{b}}_{c,k^{\prime}}^{(n)})^H \bar{\mathbf{B}}_{e,k^{\prime}}\bar{\mathbf{b}}_{c,k^{\prime}}^{(n)} ] \notag\\
        &+ 2\Re\{(\bar{\mathbf{b}}_{s})^H\bar{\mathbf{B}}_{es}\bar{\mathbf{b}}_{s}^{(n)}\} -  (\bar{\mathbf{b}}_{s}^{(n)})^H\bar{\mathbf{B}}_{es}\bar{\mathbf{b}}_{s}^{(n)} +  1\geq q_{3,k},~\label{eq:cct2}\\
    %------
        %&\log_2(q_{2,k}^{(n)}) + \frac{q_{2,k} - q_{2,k}^{(n)}}{\ln q_{2,k}^{(n)}} \leq t_{2,k}.~\label{eq:cct4}
        &\log_2(q_{2,k}^{(n)}) + (q_{2,k} - q_{2,k}^{(n)})/\ln q_{2,k}^{(n)} \leq t_{2,k}.~\label{eq:cct4}
    \end{align}
\end{subequations}
%-----------

To this end, the optimization problem \text{(P2)} is formulated as the following convex optimization problem,
%---------------
\begin{subequations}
\begin{align}
    \text{(P3)}~&\mathop{\max_{\bar{\mathbf{b}}_{c,k}, \bar{\mathbf{b}}_s, \mathbf{t}_k, \mathbf{q}_k}}~\gamma \\
\text{s.t.}~&\eqref{const0},\eqref{const6},\eqref{const10}, \eqref{const1},\eqref{const7},\eqref{eq:cct1},\eqref{eq:cct2}-\eqref{eq:cct4}.       
\end{align}    
\end{subequations}
%------------------
% which can be efficiently solved using tools like CVX~\cite{cvx}. 
%\com{Complexity analysis of the optimization problem???}.

Problem (P3) is a convex quadratically constrained quadratic program (QCQP) solved in CVX via an interior-point method. 
With $n_{\text{var}}\!\approx\!2\big(\sum_{k=1}^{K}L_{c,k}N + N\big)$ (dominated by $\{\bar{\mathbf{b}}_{c,k}\}$ and $\bar{\mathbf{b}}_s$), 
the per-iteration cost is $\mathcal{O}(n_{\text{var}}^3)$.%, so the overall complexity grows cubically with $N$, $\sum_k L_{c,k}$, and $K$.

\begin{figure}[t]
	\centering
	\vspace{0em}
	\includegraphics[width=0.40\textwidth]{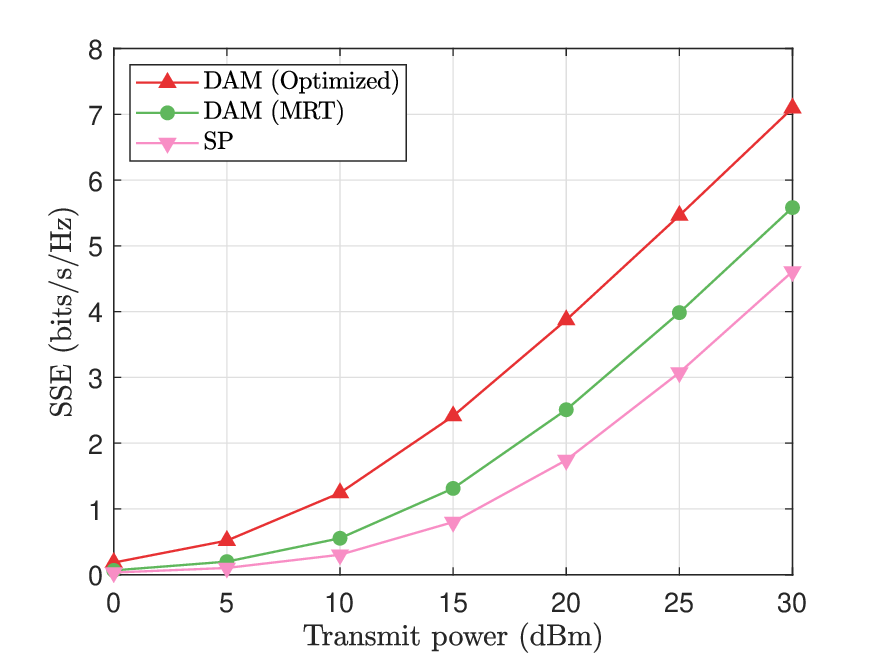}
	\vspace{-0.8em}
	\caption{ SSE comparison between DAM and SP for the worst-case UE. }
	\vspace{0.5em}
	\label{fig:SR_Power}
\end{figure}
%================================================================
%--------------------%--------------------
\vspace{-0.3em}
\section{Numerical Results} \label{sec: Numerical Results}
%--------------------
Simulation results are presented to evaluate the performance of the proposed DAM design. The BS is equipped with $N = 30$ antennas and positioned parallel to the $y$ axis. The carrier frequency is set to $f_c = 28$~GHz. The noise power spectral density for all receivers is assumed to be $-174$~dBm/Hz with a $9$~dB noise figure. The system bandwidth is $B = 128$~MHz. For a link of type $u\in\{\mathrm{L},\mathrm{N}\}$ (LoS/NLoS), the large-scale channel gain is $\alpha_u(d_m) = K_u(\frac{d_m}{d_0})^{-\epsilon_u}$, where $m$ indexes a node (UE, Eve), $d_m$ is the distance between node $m$ and the BS, $d_0=1$\,m is the reference distance, and $\epsilon_u$ is the path-loss exponent~\cite{Wang2016}. The term $K_u$ is related to the 1\,m path loss in dB by $K_u = 10^{-\beta_u^{(\mathrm{dB})}/10}$, 
with $\beta_{\mathrm{L}}^{(\mathrm{dB})}=61.4$\,dB and $\beta_{\mathrm{N}}^{(\mathrm{dB})}=72$\,dB. We set the exponents to $\epsilon_{\mathrm{L}}=2$ and $\epsilon_{\mathrm{N}}=2.92$~\cite{Wang2016}.

Figure~\ref{fig:SR_Power} plots the worst-UE SSE versus the transmit power. As expected, increasing the transmit power raises the SNR and thereby improves the SSE. Compared with the \textbf{SP} baseline, \textbf{DAM (MRT)} achieves higher SR because DAM exploits all multipath components, whereas \textbf{SP} leverages only the strongest path (\textbf{SP} aligns to the dominant path and discards the rest~\cite{Lu_DDAM}). Moreover, \textbf{DAM (Optimized)} outperforms \textbf{DAM (MRT)} by refining precoding beyond MRT, yielding additional secrecy gains. Here, MRT maximizes the desired-signal power toward the intended user but does not explicitly suppress ISI/IUI or sensing interference, while our optimized design balances power focusing with interference control.

\begin{figure}[t]
	\centering
	\vspace{0em}
	\includegraphics[width=0.40\textwidth]{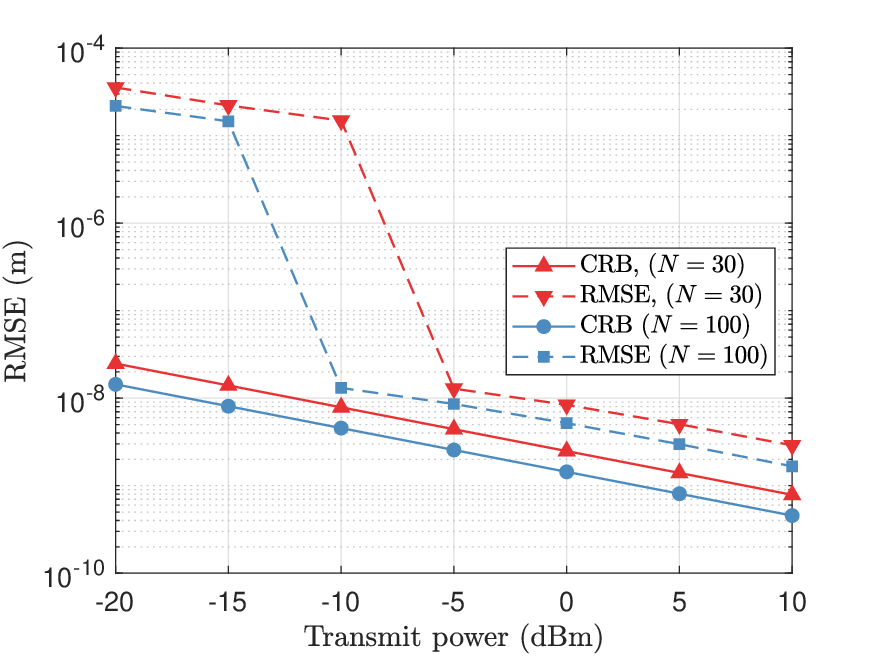}
	\vspace{-0.8em}
	\caption{ CRB versus the transmit power.}
	\vspace{0.5em}
	\label{fig:CRLB_Power}
\end{figure}

Figure~\ref{fig:CRLB_Power} shows the CRB and the RMSE of the proposed delay estimator versus the transmit power. The RMSE obtained by the maximum likelihood estimator approaches the CRB, indicating that the bound is tight and thus informative for performance prediction. 
We compare the CRB for $N=30$ and $N=100$; the CRB decreases with larger $N$ since a higher array gain improves Fisher information. This confirms that increasing array aperture enhances sensing precision, and the proposed estimator exploits array gain for accurate delay estimation.

%====================================================
%================================================================

%Figure~\ref{fig:SR_CRLB} plots the SR versus the CRB threshold $\Gamma$. The SR increases with $\Gamma$ because a looser sensing-accuracy requirement allows the transmitter to place less emphasis on matching the target 
%direction and allocate more precoding gain to the communication links, thereby raising the UEs’ SNR and, in turn, the SR. Moreover, the SR improves as the antenna count $N$ increases, since a larger array provides higher array gain, which makes DAM more effective at mitigating ISI/IUI and strengthening the desired signal.

%\begin{figure}[t!]
%  \centering
%  \includegraphics[width=3.4in]{figure3.pdf}
%  \caption{SR versus CRLB.}
%  \label{fig:SR_CRLB}
%\end{figure}

\vspace{-0.1em}
\section{Conclusion}
This paper introduced DAM for secure ISAC systems. DAM aligns multipath components at intended users, while causing delay and angle mismatches at the Eve, enhancing security without extra power. We derived the SSE and CRB, and then conceived a two-stage sensing protocol. A path-based ZF precoding scheme was also proposed to maximize the minimum SSE under CRB and power constraints. Simulations verified that DAM outperforms the SP benchmark and achieves superior secrecy performance with larger antenna arrays.

\vspace{-0.1em}
\bibliographystyle{IEEEtran}
\bibliography{IEEEabrv,PLS_ISAC}
\end{document}